\documentclass[aps,preprintnumbers,nofootinbib,superscriptaddress]{revtex4}

\usepackage{graphicx}
\DeclareGraphicsExtensions{.pdf}

\usepackage{amsfonts,amssymb,amsmath}
\usepackage{amsthm}

\newcommand{\comment}[1]{}
\newcommand{\ket}[1]{\left |  #1 \right\rangle}

\theoremstyle{plain}

\theoremstyle{definition}

\begin{document}

\title{Unconditionally Secure Bit Commitment by Transmitting
  Measurement Outcomes}

\author{Adrian \surname{Kent}}
\affiliation{Centre for Quantum Information and Foundations, DAMTP, Centre for
  Mathematical Sciences, University of Cambridge, Wilberforce Road,
  Cambridge, CB3 0WA, U.K.}
\affiliation{Perimeter Institute for Theoretical Physics, 31 Caroline Street North, Waterloo, ON N2L 2Y5, Canada.}

\date{ August 2011; revised April 2012}

\begin{abstract}
We propose a new unconditionally secure bit commitment scheme
based on Minkowski causality and the properties of quantum information. 
The receiving party sends a number of 
randomly chosen BB84 qubits to the committer at a given point in space-time.
The committer carries out measurements in one of the two
BB84 bases, depending on the committed bit value,
and transmits the outcomes securely at (or near) 
light speed in opposite directions to 
remote agents.  
These agents unveil the bit by returning the outcomes to adjacent 
agents of the receiver. 
The protocol's security relies only on simple properties of quantum information 
and the impossibility of superluminal signalling. 
\end{abstract}

\maketitle

{ \bf Introduction} \qquad Much research on physics and cryptography has been
devoted to the problem of bit commitment, a basic cryptographic task
which also has many applications (e.g. \cite{broadbenttapp}).
In a bit commitment protocol, the committer, Alice, carries
out actions that commit her to a particular bit
value (or, in the quantum case, a particular weighted superposition
of bit values).  She can later (or, in the relativistic case, at one or 
more points in the causal future of the commitment), if she chooses, 
give the receiver, Bob, classical or quantum information that
unveils the committed bit.   Ideally, Bob should have an absolute
guaranteee that Alice is committed by her initial actions, and Alice
should have an absolute guarantee that Bob can learn no information
about the committed bit before she unveils.

Beside its intrinsic crytographic importance and its use 
in applications, bit commitment also has intriguingly
deep connections to fundamental physics. 
Initially, work in this area 
focussed entirely on bit commitment protocols based on
non-relativistic quantum mechanics.  Bennett and Brassard invented the
first quantum bit commitment protocol \cite{BBeightyfour} and showed
that it is secure against both parties given current technology, but
insecure if Alice has a quantum memory.  Later attempts at
unconditionally secure non-relativistic protocols (e.g. \cite{BCJL})
were ultimately shown to be futile by celebrated results of Mayers
\cite{mayersprl,mayersone}, Lo and Chau \cite{lochauprl,lochau}, later
further extended \cite{mkp,darianoetal}, showing that no
unconditionally secure non-relativistic quantum bit commitment
protocol exists.

However, the world is relativistic, and in particular to good approximation 
space-time in the Earth's local neighbourhood is Minkowski. 
Allowing relativistic protocols, which exploit 
the signalling constraints implied 
by Minkowski causality, radically changes the picture.   
It was shown some time ago that classical relativistic protocols
{\it can} evade the Mayers and Lo-Chau no-go theorems
 \cite{kentrel,kentrelfinite}.  These schemes are 
provably secure against all classical attacks and
against Mayers and Lo-Chau's quantum attacks, and are conjectured
to be unconditionally secure, provided that quantum theory is
correct and the background space-time is approximately Minkowski.
Since small corrections due to general relativity can be allowed
for, and the possibility of an adversary
surreptitiously making major changes to the local
space-time geometry is beyond any presently
forseeable science and technology, such security appears
robust for the forseeable future.

Recently, a new quantum cryptographic idea was 
introduced\cite{nosummoning} and
applied to bit commitment \cite{bcsummoning}.  
The idea is that the committer is
required to send a quantum state, supplied by and known to the
receiver but unknown to the committer, at light speed over secure
quantum channels in one of two or more directions.  This gives a
practical bit commitment scheme which is also easily shown to be 
unconditionally secure against general attacks 
\cite{bcsummoning}.  It relies essentially on the control over
physical information that special relativity and quantum theory
together allow -- specifically, on the no-summoning theorem
\cite{nosummoning}, which also has other cryptographic applications
\cite{otsummoning}.

Here we propose another new quantum relativistic bit commitment
protocol.  It uses some similar intuitions to that of 
Ref. \cite{bcsummoning}, in that the committer is forced
by Minkowski causality to choose a particular commitment 
at the space-time point where the protocol starts.  
However, it relies on a different physical principle --
essentially, on the impossibility of completing a nonlocal measurement on
a distributed state outside the joint future light cone of
its components.    
Its implementation requires minimal quantum resources: 
the receiver needs to send quantum states (which can be 
unentangled qubits) to the committer, who needs to carry out
individual measurements on them as soon as they are received. 
No further quantum commmunication is required by either party;
nor do they require any entanglement, collective measurements, or  
quantum state storage. 

As usual in quantum cryptography, we present the protocol in an 
idealized form assuming perfect state preparations, transmissions
and measurements.  This poses no significant 
issue of principle here: the
protocol clearly remains secure in the
presence of errors up to a certain threshold.  

We also make idealizations about the relativistic geometry and 
signalling speed, supposing that Alice and Bob each have agents in secure
laboratories infinitesimally separated from the points $P$, $Q_0$ and
$Q_1$, Alice can signal at precisely light speed, and all information
processing is instantaneous.  
These too pose no problem of principle.  Like other protocols of
this type \cite{nosummoning,bcsummoning}, the protocol remains 
secure in realistic implementations with finite separations and 
near light speed communication.  If these corrections are
small, the only significant effect is that 
Bob is guaranteed that Alice's commitment
is binding from some point $P'$ in the near causal future of $P$,
rather than from $P$ itself \cite{bcsummoning}.  

{\bf Bit commitment based on transmitting quantum measurement
  outcomes} \qquad We give a simple version of the scheme using qubit states 
and measurements in the standard BB84 basis \cite{BBeightyfour}.
The scheme obviously generalizes to other sets of qubit states and
measurements, to qudits, and to variants with more or differently
located unveiling points.

Alice and Bob agree on a space-time point $P$, a set of coordinates
$(x,y,z,t)$ for Minkowski 
space, with $P$ as the origin, and (in the simplest case)
two points $Q_0 = (x,0,0,x)$  and $Q_1 = (-x,0,0,x)$ light-like  
separated from $P$. 
They each have agents, separated in secure laboratories, adjacent to 
each of the points $P$, $Q_0$, $Q_1$.  To simplify for the moment,
we take the distances from the labs to the relevant points as
negligible.    

Bob securely preprepares a set of qubits $\ket{\psi_i}_{i=1}^N$ independently
randomly chosen from the BB84 states $\{ \ket{0} , \ket{1} , \ket{+},
\ket{-} \}$ (where $\ket{\pm} = \frac{1}{\sqrt{2}} ( \ket{0} \pm
\ket{1} )$) and sends them to Alice to arrive (essentially) at $P$. 
To commit to the bit value $0$, Alice measures each state in the $\{ \ket{0} ,
\ket{1} \}$ basis, and sends the outcomes over secure classical 
channels to her agents at $Q_0$ and $Q_1$. 
To commit to $1$, Alice measures each state in the $\{ \ket{+} ,
\ket{-} \}$ basis, and sends the outcomes as above. 
Alice's secure classical channels could, for example, be created
by pre-sharing one-time pads between her agent at $P$ and those
at $Q_0$ and $Q_1$ and sending pad-encrypted classical signals.
If necessary or desired, these pads could be periodically replenished by 
quantum key distribution links between the relevant agents. 

To unveil her committed bit, Alice's agents at $Q_0$ and $Q_1$
reveal the measurement outcomes to Bob's agents there. 
After comparing the revealed data to check that the declared
outcomes on both wings are the same (somewhere in the intersection
of the future light cones of $Q_0$ and $Q_1$), and that both are 
consistent with the list of  states sent at $P$, Bob accepts the commitment
and unveiling as genuine.  If the declared outcomes are different,
Bob has detected Alice cheating. 

{\bf Security} \qquad The 
protocol is evidently secure against Bob, who learns nothing
about Alice's actions until (if) she chooses to unveil the bit. 

Alice is constrained in that she has to be able to
reveal her commitment data at both $Q_0$ and
$Q_1$, since Bob's agents at these points verify the 
timing and location of the unveilings, and then later
compare the data to check they are consistent. 
By Minkowski causality, Alice's ability to
unveil data consistent with a $0$ or $1$ commitment at $Q_0$ depends
only on operations she carries out on the line $P Q_0$.  Suppose that
she has a strategy in which she carries out some operations at $P$,
but these leave her significantly uncommitted, in the sense that her
optimal strategies $S_i$ for successfully unveiling the bit values
$i$, by carrying out suitable operations in the causal future of $P$,
have success probabilities $p_i$, with $p_0 + p_1 > 1 + \delta$, for
some $\delta >0$.  By Minkowksi causality, any operations she carries
out on the half-open line segment $\left( P , Q_0 \right]$ cannot
affect the probability of producing data at $Q_1$ consistent with a
succesful unveiling of either bit value $i$ there.  In particular, if
she follows the instructions of strategy $S_0$ on $\left( P , Q_0
\right]$, and the instructions of strategy $S_1$ on $\left( P , Q_1
\right]$, she has probabilities $p_i$ of producing data consistent
with a successful unveiling of bit value $i$ at $Q_i$, and hence
probability at least $\delta$ of producing data consistent with a
successful unveiling of bit value $0$ at $Q_0$ {\it and} with a
successful unveiling of bit value $1$ at $Q_1$.  

This means that, with
probability at least $\delta$, by combining her data at $Q_0$ and $Q_1$
at some point in their joint causal future, Alice can produce data
consistent with both sets of measurements in complementary bases.  
Thus, for example, for each state $\ket{\psi_i}$, she can identify
a subset of $2$ states from $\{ \ket{0}, \ket{1}, \ket{+}, \ket{-}
\}$, one from each basis, which must include $\ket{\psi_i}$.  
By choosing the security parameter $N$ large
enough, Bob can ensure Alice's overall success probability, for
correctly identifying such a subset for each of the $N$ states, is smaller
than $\delta$, for any given $\delta > 0$, i.e. that Alice has no cheating
strategy of the type described.  Hence the protocol is also
secure against Alice.

\begin{figure}[t]
\centering
\includegraphics[scale=0.5]{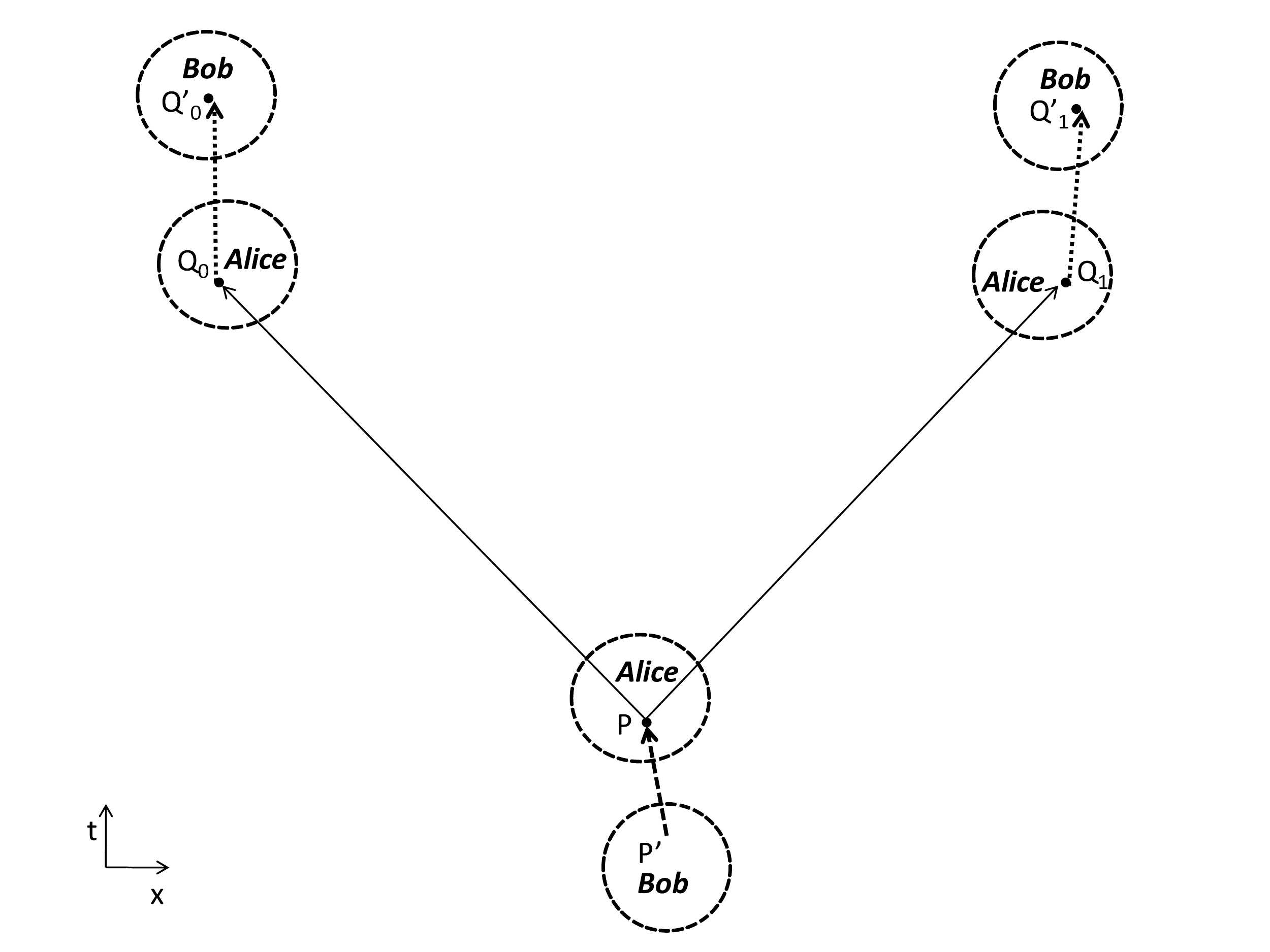}
\caption{A non-ideal implementation of the protocol in 
$1+1$
dimensions.   (Not to scale.)  Alice and Bob control disjoint regions of space-time,
representing their respective secure laboratories.   Bob generates
random BB84 quantum states at $P'$ and sends them (dashed arrow) to 
Alice at point $P$, where she measures them in
her chosen basis.   She reports the results (solid arrows) via secure classical
(near) light-speed channels to her agents at the points $Q_i$, 
who relay them (dotted arrows) to Bob's agents at the nearby points $Q'_i$.} 
\end{figure}

{ \bf Discussion} \qquad We now have a variety
of practical schemes for bit commmitment -- the one described above and those
of Refs. \cite{kentrelfinite,bcsummoning} -- including 
two distinct intrinsically quantum and practical schemes with complete
security proofs, solving a problem once thought unsolvable.
The interesting open questions now seem to be understanding 
the full variety of techniques and schemes, comparing the
resources they require, and identifying the most practical
in any given context.   

The protocol given here has some significant practical advantages. 
Unlike the protocol of Ref. \cite{kentrelfinite}, it does not
require Alice to preshare data and coordinate her commitment 
timings in advance between agents at space-like separated points.  
Compared to the protocol of Ref. \cite{bcsummoning}, it requires
significantly less advanced quantum technology: 
it needs reasonably reliable preparation (by Bob)
and measurement (by Alice) of single qubits, but no entanglement, no collective
measurements, and no secure quantum communication channels for either
party.  Note also that the protocol remains secure even if Alice has
a low detection efficiency for the transmitted qubits, so long as (i) she
can carry out reasonably reliable measurements on those she does
detect, and (ii) she can report quickly back to Bob (i.e. at or close
to $P$) which qubits were detected and measured.  

Of course, Bob needs to be able to generate randomly chosen quantum
states securely in his lab, but once they are generated it does not
matter significantly if Alice is able to carry out quantum operations
on them immediately: the only effect is to slightly alter the
space-time region in which Bob is confident Alice is committed.  
Alice also requires secure classical communication channels: these could  
be implemented using sufficiently long preshared one time pads, or by
shorter preshared pads expanded indefinitely using quantum key 
distribution.

As noted above, in a 
realistic implementation, Alice and Bob's agents would have
nonzero separation from the space coordinate of $P$, and likewise
from $Q_0$ and $Q_1$.  This affects the discussion only in that
the causal geometry is a little more complex.\cite{bcsummoning}   
For example, Bob cannot guarantee that Alice was committed at any point in 
in the interior of the intersections of the past light cones
of the points $Q^B_0$ and $Q^B_1$ where he actually receives
her unveiling data.  

Like the protocol of Ref. \cite{bcsummoning}, 
our protocol is immune to a Mayers-Lo-Chau attack
\cite{mayersprl,lochauprl} , and for similar reasons.
Mayers-Lo-Chau's arguments correctly imply that there is 
a unitary operation on a spacelike hyperplane through $Q_0$ 
and $Q_1$ that, mathematically, maps quantum data consistent
with a $0$ commitment and unveiling to data consistent with
a $1$, and vice versa.   However, this operation cannot be 
implemented physically on the hyperplane: doing so would
violate Minkowski causality.  

A realistic implementation must also allow for errors
in Bob's preparations, the communication of the states to
Alice, and Alice's measurements.   So long as these errors
are small, this makes no essential difference: Bob needs only
test that Alice's declared outcomes are statistically consistent
with the measurements corresponding to one bit value commitment,
and statistically inconsistent with the other.  

Note that, like all technologically unconstrained quantum bit commitment
protocols\cite{kentbccc,kentshort}, the protocol here does not 
prevent Alice from committing to a quantum superposition of 
bits.   She can simply input a superposition $\alpha \ket{0} +
\beta \ket{1}$ into a quantum computer programmed to implement
the two relevant quantum measurement interactions for inputs 
$\ket{0}$ and $\ket{1}$ and to send two copies of the quantum
outcome data towards $Q_0$ and $Q_1$, and keep all the data at
the quantum level until (if) she chooses to unveil.   
This gives her no advantage in stand-alone applications of bit
commitment, for example for making a secret prediction: it 
does, however, mean that one cannot assume that in a task
involving bit commitment subprotocols, any unopened
bit commitments necessarily had definite classical bit values,
even if all unveiled bit commitments produced valid classical
unveilings. 

As with the protocols of Ref. \cite{kentrelfinite,bcsummoning}, the present
protocols can be chained together in sequence, allowing longer term
bit commitments and flexibility in the relation between the commitment
and unveiling sites (in particular, they need not be lightlike
separated).  Full security and efficiency analyses for these chained
protocols remain tasks for future work.   

Relativistic quantum cryptography 
allows strategies that 
make no sense in non-relativistic classical or quantum cryptography.  
For example, one intriguing and rather Zen-like feature 
that the present protocol shares with that of
Ref. \cite{bcsummoning} is that Alice makes her commitment without
giving {\it any} information, classical or quantum, to Bob. 
This departs from the usual way of thinking about bit commitment,
in which Alice commits herself by handing over data that is 
in some way encrypted, and unveils by handing over some form
of decryption key.  Instead, here, relativistic causality 
forces Alice to commit herself so that she can later 
make a valid unveiling.  She could delay her measurement choice,
but if she does she cannot get valid measurement outcomes to 
both unveiling points.   In a sense, Bob receives the commitment
and unveiling together, and the unveiling carries with it a
guaranteed record of the past commitment. 

More broadly, our picture of relativistic quantum cryptography is changing 
quite fast, from a field with some interesting niche applications
\cite{kentrelfinite, colbeckkent,kwiat} to a distinct branch of physics-based
cryptography  with a potentially very wide range of applications.
These include interesting  cryptographic tasks whose 
definition itself is intrinsically relativistic \cite{otsummoning}.
They also include unconditionally secure schemes for quantum 
tagging \cite{kenttaggingcrypto}
(also called quantum position authentication), 
conditionally secure quantum tagging 
schemes \cite{taggingpatent,malaney,buhrmanetal,kms}
based on slightly weaker security
assumptions, and more generally, a large class of schemes for more 
general tasks in position-based quantum cryptography \cite{buhrmanetal}.  
We hope all these developments will stimulate further interest in 
the theory and practical implementation of relativistic
quantum protocols.


\acknowledgments

I am grateful to Serge Massar and Jonathan Silman for helpful discussions.
This work was partially supported by an FQXi mini-grant and by
Perimeter Institute for Theoretical Physics. Research at Perimeter
Institute is supported by the Government of Canada through Industry
Canada and by the Province of Ontario through the Ministry of Research
and Innovation.


\end{document}